\documentclass[twocolumn,groupedaddress,showpacs,floatfix,notitlepage,color]{revtex4-1}
\usepackage{graphicx}
\usepackage{epsfig}
\usepackage{amsfonts}
\usepackage{color}
\usepackage{ulem}
\usepackage{amsmath}
\usepackage{mathrsfs}
\usepackage{subfigure}
\usepackage[autostyle]{csquotes}

\begin{document}

\title{Stochastically forced dislocation density distribution in plastic deformation}

\author{Amit K Chattopadhyay}
\affiliation{                    
Mathematics, Aston University, Aston Triangle, Birmingham, B4 7ET, UK \\
Aston Materials Centre, Aston University, Aston Triangle, Birmingham, B4 7ET, UK}
\email{a.k.chattopadhyay@aston.ac.uk}
\author{Elias C. Aifantis}
\affiliation{Laboratory of Mechanics and Materials, Aristotle University of Thessaloniki, GR-54124 Thessaloniki, Greece \\
Michigan Technological University, Houghton MI 49931, USA \\
ITMO University, St. Petersburg 197101, Russia and BUCEA, Beijing 100044, China}
\email{mom@mom.gen.auth.gr}
\begin{abstract}
The dynamical evolution of dislocations in 
plastically deformed metals is controlled by both deterministic factors arising out of applied loads and stochastic effects appearing due to fluctuations of internal stress. Such type of stochastic dislocation processes and the associated spatially inhomogeneous modes lead to randomness in the observed deformation structure. 
Previous studies have analyzed the role of randomness in such textural evolution but none of these models have considered the impact of a finite decay time (all previous models assumed instantaneous relaxation which is \enquote{unphysical}) of the stochastic perturbations in the overall dynamics of the system. The present article bridges this knowledge gap by introducing a colored noise in the form of an Ornstein-Uhlenbeck noise in the analysis of a class of linear and nonlinear Wiener and Ornstein-Uhlenbeck processes that these structural dislocation dynamics could be mapped on to. Based on an analysis of the relevant Fokker-Planck model, our results show that linear Wiener processes remain unaffected by the second time scale in the problem but all nonlinear processes, both Wiener type and Ornstein-Uhlenbeck type, scale as a function of the noise decay time $\tau$. The results are expected to ramify existing experimental observations and inspire new numerical and laboratory tests to gain further insight into the competition between deterministic and random effects in modeling plastically deformed samples.
\end{abstract}
\date{\today}


\maketitle

\section{Introduction}

Inhomogeneities in the evolution of dislocation density and other nano/microstructures can be modeled by introducing elements of stochasticity into the constitutive equations through a set of internal variables. An interesting approach to introduce stochastic effects in plastic deformation and model the randomized contours of the dislocation propagation fronts was advanced in  \cite{haehner1996}, followed by a related work in  \cite{avlonitis2000}, as well as in  \cite{zaiser2003,zaiser2006}. The latter set of stochastic models considered strong heterogeneity of the real samples, in line with the well established {\it gradient models} \cite{aifantis1984, aifantis1992, aifantis1995,aifantis2011,morris2011} that analyzed the emergence of deterministic spatial ideformation patterns. In the context of continuum modeling, these gradients can be mapped on to the equivalent of complete \enquote{dynamical balance laws} for the internal variables, containing both a rate and a flux term which account for the transport of microstructures through elementary material volume. The problem then becomes one of multiple length scales and relates to the effective description of the dependence of the evolution dynamics of the dominant microstructure on these length scales. One such paradigmatic model is the Walgraef-Aifantis model \cite{walgraef1985} in which diffusion-like relaxation of the dislocation dynamics was considered on spatially periodic dislocation patterns in cyclic plastic deformation. 

A stochastic approach to plastic deformation incorporating the omnipresent micro-nanostructure heterogeneity is based on the introduction of an explicit stochastic forcing term in the set of internal variable constitutive equations.
This has recently been adopted through the introduction of an explicit stochastic forcing term in the set of internal variable constitutive equations \cite{haehner1996,avlonitis2000}. The underlying statistical mechanics based approach, in effect a Langevin formulation \cite{risken}, of studying heterogeneous plastic deformation was pioneered in \cite{haehner1996}; an extension of this approach was later used in modeling unidirectional plastic deformation in \cite{avlonitis2000}. The origin of the stochasticity was attributed to \enquote{slip lines} or \enquote{slip bands} developed during shearing. Both for white and quenched noise, this leads to randomness in the slip profile, as also reported in previous experimental studies \cite{neuhauser1983,morris2011}. In line with this initial work \cite{haehner1996}, follow-up studies successfully analyzed dislocation and slip-channel patterning \cite{zaiser1997,zaiser1996}, as well as dislocation clustering in cyclic plastic deformation \cite{haehner1996appl,bako1999}, including the impact of stochastic nano-scale inelastic events in amorphous media \cite{yang2014}.

Almost all such models, though, relied on a stationary state distribution description for the dislocation density profile and, hence, were not able to exactly analyze the dynamical evolution of such patterns. Using an exactly solvable real time dependent model, the first major non-stationary state stochastic model was analyzed by Avlonitis, et al \cite{avlonitis2000}. The authors were able to completely characterize the evolution of the distribution function under the assumption of a stochastic white noise, that excludes the possibility of a finite relaxation time for the stochastic perturbation to survive. In real dynamical terms, such effects are often known to be a vital component of the dynamics \cite{murray} and there is no reason to assume that the same is not true here as well. A consummate summary of such efforts can be availed from the review by Aifantis \cite{aifantis2014}.

This article propounds a complementary approach to incorporate finite time relaxation due to inhomogeneous perturbations through the incorporation of a finite decay time $\tau$. This finite relaxation time of stochastic perturbation is introduced through an Ornstein-Uhlenbeck noise. The emphasis here is on studying the qualitative and quantitative changes that such a non-trivial $\tau$ will bring on the Avlonitis-Zaiser-Aifantis (AZA) model \cite{avlonitis2000}. Section \ref{Langevin} will be a general reminder of the AZA model, leading to the formulation of the Langevin dynamics, that now will be driven by a stochastic Ornstein-Uhlenbeck noise. Section \ref{FP} will derive a Fokker-Planck model\cite{risken}, starting from the Langevin description in Section 2. This Fokker-Planck model will then be solved for all times $t$ exactly and studied for the dependence of the decay time $\tau$ in its dynamics. Section \ref{FP} has been subdivided into four subsections, each discussing and comparing respectively the Wiener process (subsection \ref{wiener}), Ornstein-Uhlenbeck process (subsection \ref{ornstein_uhlenbeck}), generalized Ornstein-Uhlenbeck process (subsection \ref{general_ornstein_uhlenbeck}) and the generalized exponential process (subsection \ref{general_exponential}), as was previously done in \cite{avlonitis2000} for instantaneous relaxation but now studied in presence of a non-trivial Ornstein-Uhlenbeck time scale $\tau$. The conclusion Section \ref{conclusion} will summarize the impact of a time-dependent colored noise in the temporal dynamical patterning of plastic deformation based on comparison with the previously discussed (in subsections \ref{wiener}-\ref{general_exponential}) four subclasses.

\section{Langevin dynamics of internal variables during plastic deformation}
\label{Langevin}
In line with \cite{avlonitis2000}, our starting premise will be a set of evolving internal variables $\{\phi_i\}$ whose time evolution will define the course of plastic deformation to be governed by plastic straining only. Any other process that influences the microstructural evolution (diffusional processes, static/dynamic recovery, etc.) may be neglected. In this case, all terms in the evolution equations scale as the strain rate

\begin{equation}
\partial_t \phi_i = F_i(\{\phi_i\},\{q_k\})\:{\dot \gamma} \to \partial_\gamma \phi_i = F_i(\{\phi_i\},\{q_k\}),
\label{phi_Equation}
\end{equation}

\noindent
where $\{q_k\}$ denote a set of external control parameters and $\gamma$ represents the equivalent plastic shear strain. The detailed study of these class of constitutive models can be found in \cite{avlonitis2000}; we reiterate portions of this representation for the sake of continuity. The above equation can be reconsidered as a stochastic equation of evolution by taking into account fluctuations of the shear strain rate around its mean value

\begin{equation}
\dot \gamma = <\dot \gamma> +\: \delta \dot \gamma.
\label{gamma_Equation}
\end{equation}

In the above formulation, the quantity \enquote{$<>$} represents an \enquote{ensemble average} over all noise realization, details of which will be shortly presented. The amplitude of these fluctuations can be related to the average strain rate $<\dot \gamma>$, the mean internal (back) stress $f_{\text{int}}$ experienced by the ensemble of moving dislocations, and the strain-rate sensitivity S of the material as follows

\begin{equation}
\frac{\delta {\dot \gamma}^2}{{<\delta {\dot \gamma}>}^2} = \frac{<\delta v^2>}{{<v>}^2}=\frac{<f_{\text{int}}>}{S}.
\label{sensitivity_Equation}
\end{equation}

In the above equation, $v$ represents the dislocation velocity through the following relation: $\dot \gamma=\rho_m\:b\:v$, $\rho_m$ being the density of mobile dislocations and $b$ is their Burgers vector. This leads to the perturbative description $v=<v>+\:\delta v$, where $<v>$ is the mean component of the dislocation velocity with $\delta v$ being the fluctuation part ($<\delta v >=0$) of $v$ which increases closer to the sheared layers. The fluctuation correlation time $t_{\text{corr}}$ that defines the mean collective dislocation time is associated with the mean activity time of the slip line and can be quantified as

\begin{equation}
t_{\text{corr}} = \frac{\Gamma_{\text{corr}}}{<\dot \gamma>} = \frac{\rho_m bL}{<\dot \gamma>}=\frac{L}{<v>},
\label{time_corr_Equation}
\end{equation}

\noindent
where $\Gamma$ is the mean macroscopic shear strain and $L$ is a typical slip-line length.

Without any loss of generality, we generalize the remit of equation (\ref{phi_Equation}) for all variables (the suffix \enquote{i} is removed henceforth) to describe the microstructural evolution in terms of a single variable $\phi$ as follows

\begin{equation}
\partial_t \phi = F(\phi)\:{\dot \gamma} \to \partial_t \phi= F(\phi)<\dot \gamma> + F(\phi) \:\delta \dot \gamma.
\label{general_phi_Equation}
\end{equation}

Introducing the average macroscopic strain $\Gamma$ as the equivalent of the time variable and with further rescaling, we have $d\Gamma = <\dot \gamma>\:dt$, leading to

\begin{equation}
\partial_\Gamma \phi = F(\phi)+F(\phi) \:\frac{\delta \dot \gamma}{<\dot \gamma>}.
\label{rescaled_Equation}
\end{equation}

In line with the AZA-approach \cite{avlonitis2000}, the strain-rate fluctuations $\frac{\delta \dot \gamma}{<\dot \gamma>}$ are idealized by a standard correlated stochastic process driven by a noise $Q\eta(\Gamma)$ whose amplitude (squared) is given by

\begin{subequations}
\begin{equation}
Q^2 = \frac{<\delta \dot \gamma^2>}{{<\dot \gamma>}^2}\:\Gamma_{\text{corr}} = {Q_0}^2\:H^2(\phi),
\end{equation}
\begin{equation}
<\eta(\Gamma) \eta(\Gamma')>=\left( \frac{D_0}{\tau}\right)\:\exp\left(-\frac{|\Gamma-\Gamma'|}{\tau}\right).
\end{equation}
\label{noise_Equation}
\end{subequations}

In the above, we have introduced an Ornstein-Uhlenbeck noise $\eta(\Gamma)$ \cite{fox1986}, that decays at the rate of $\tau$, to account for a second finite timescale in the problem. This time scale represents the competition between the stochastic and deterministic components of the dynamics, that for all practical purposes, could be estimated as a number comparable to the relaxation time $\tau_{\text{on}}$ of the system. For $\tau>\tau_{\text{on}}$, the decay time will assume prominence and renormalize the effective relaxation rate of the system. Equation (\ref{general_phi_Equation}) then becomes

\begin{equation}
\partial_\Gamma \phi = F(\phi) + Q_0 \:\left( F(\phi) H(\phi)\right)\:{\eta}.
\label{new_phi_Equation}
\end{equation}

Using $G(\phi) = F(\phi)\:H(\phi)$, the above equation reduces to

\begin{equation}
\partial_\Gamma \phi = F(\phi) + Q_0 \:G(\phi) \:{\eta}.
\label{phi_gamma_Equation}
\end{equation}

As in \cite{avlonitis2000}, we now resort to a non-linear transformation of variables abiding the definition

\begin{equation}
u(\phi,\phi') = \displaystyle \int_{\phi'}^{\phi}\:\frac{d{\tilde \phi}}{G(\tilde \phi)}.
\label{u_Equation}
\end{equation}

In terms of the new variable $u$, we can now rewrite Eq. (\ref{new_phi_Equation}) to arrive at our Langevin model as follows

\begin{equation}
\frac{\partial u}{\partial \Gamma} = \frac{F(u)}{G(u)} + Q_0 {\eta}.
\label{Langevin_Equation}
\end{equation}

Eq. (\ref{Langevin_Equation}) above is the Ornstein-Uhlenbeck noise driven Langevin dynamics for the dislocation pattern evolution and our starting model equation. The model has two time scales, an inherent relaxation rate defined by the deterministic dynamics which now competes against the relaxation rate of noise, represented by the decay time $\tau$. In the following section, we derive the Fokker-Planck version of the Langevin dynamics starting from our ground model.

\section{Fokker-Planck dynamics of plastic deformation variables}
\label{FP}

To arrive at a form for the time variation of the probability density function (PDF) of the Langevin dynamics represented by the variable $u(\Gamma)$ in Eq. (\ref{Langevin_Equation}) of the previous section, we follow the Fox module \cite{fox1986}. For a Langevin model represented as $\dot X(t) = W(X) + g(X) \eta(t)$, one can show, following some algebra, that the corresponding Fokker-Planck model representing the time evolution of the PDF $P(X,t)$ of the variable $X(t)$ is given by

\begin{eqnarray}
\frac{\partial P}{\partial t}& =& -\frac{\partial}{\partial X}\big[W(X)\:P\big]\: +\:\frac{1}{2}\frac{\partial}{\partial X}g(X) \nonumber \\
&\times& \frac{\partial}{\partial X}\bigg[\frac{g(X)}{1-\tau W'(X)}-\frac{\tau W(X) g'(X)}{{[1-\tau W'(X)]}^2}\bigg]P,
\label{fokkerplanck_Equation}
\end{eqnarray}
where $<\eta(t)\eta(t')>=\frac{1}{\tau}\exp\big(-\frac{|t-t'|}{\tau}\big)$. Expanding the $\tau$-dependent denominator to the first leading order in $\tau$, we obtain

\begin{eqnarray}
\frac{\partial P}{\partial t} &=& -\frac{\partial}{\partial X}\bigg[W(X)\:P\bigg]\:+\:\frac{1}{2}\frac{\partial}{\partial X}g(X) \\ \nonumber 
&\times &\frac{\partial}{\partial X}\bigg[ g(X) + \tau g(X) W'(X)-\tau W(X) g'(X) \bigg]P. 
\label{fp_Equation}
\end{eqnarray}

For our purpose, $W(u)=\frac{F(u)}{G(u)}$ and $g(u)=Q_0$. This approach can be further generalized to study the full Ornstein-Uhlenbeck colored noise spectrum as could be defined through a $u$-dependent $Q_0(u)$. For simplicity, we will assume $Q_0$ = constant. Defining the PDF as $P(u,t)={<\delta(u-u(t))>}_\eta$, where $\delta$ is the Dirac-Delta function and the average \enquote{$<>$} is over all noise realization $\eta$, the Fokker-Planck version of our Langevin model can be derived using the approach detailed in \cite{fox1986}, which then is given by

\begin{eqnarray}
\frac{\partial P}{\partial \Gamma} &=& -\frac{\partial}{\partial u}\bigg[\frac{F(u)}{G(u)}\:P\bigg]\nonumber \\
&+&\:\frac{{Q_0}^2}{2}\frac{\partial^2}{\partial u^2}\bigg[ 1+ \tau \frac{\partial}{\partial u}\left( \frac{F(u)}{G(u)} \right) \bigg]P.
\label{model_fp_Equation}
\end{eqnarray}

In what follows, we will separately consider the following four cases:
\begin{enumerate}
\item[A)] $\frac{F(u)}{G(u)}=1$; this is the Wiener process;
\item[B)] $\frac{F(u)}{G(u)}=u$; this is the Ornstein-Uhlenbeck process;
\item[C)] $\frac{F(u)}{G(u)}=u^n$, where $n>1$; this is the {\it generalized Ornstein-Uhlenbeck} process;
\item[D)] $\frac{F(u)}{G(u)}=\exp(\alpha u)$ ($\alpha<0$), where $-\frac{1}{\alpha}$ is any decay constant characterizing the dynamical process; this is the {\it generalized exponential} process.
\end{enumerate}

\subsection{Wiener Process}
\label{wiener}

This process will be trivial for the case of a colored Ornstein-Uhlenbeck noise. This is because, since $\frac{F(u)}{G(u)}=1$, the $\tau$-dependent term in Eq. (\ref{model_fp_Equation}) will vanish, resulting in an unchanged Wiener process as was previously reported in \cite{avlonitis2000}.  In other words, starting from the following Fokker-Planck equation

\begin{equation}
\frac{\partial P}{\partial \Gamma}= -\frac{\partial P}{\partial u}+\:\frac{{Q_0}^2}{2}\frac{\partial^2 P}{\partial u^2},
\label{wiener_fp_Equation}
\end{equation}

\noindent
which then leads to a Gaussian solution $P(u,\Gamma) = \frac{1}{\sqrt{2\pi {Q_0}^2 \Gamma}}\:\exp\left(-\frac{{(u-\Gamma)}^2}{2{Q_0}^2 \Gamma}\right)$ for the initial condition $P(u,0)=\delta(u)$.
 
\subsection{Ornstein-Uhlenbeck Process}
\label{ornstein_uhlenbeck}

In this case, the relevant Fokker-Planck equation has the representation

\begin{equation}
\frac{\partial P}{\partial \Gamma}= -\frac{\partial}{\partial u}\big(uP\big)+\:\frac{{Q_0}^2}{2}\frac{\partial^2}{\partial u^2}\big[(1+\tau)P\big].
\label{ou_fp_Equation}
\end{equation}

This leads to the following exact time-dependent solution for the PDF

\begin{equation}
P(u,\Gamma) =  \frac{\exp\left(-\frac{{u}^2}{{Q_0}^2(1+\tau)(1-e^{-2\Gamma})}\right)}{\sqrt{2\pi {Q_0}^2 (1+\tau) (1-e^{-2\Gamma})}}.
\label{ou_fp_soln}
\end{equation}

This can be simplified as follows

\begin{equation}
P(u,\Gamma) =  \frac{\exp\left(-\frac{{u}^2}{{Q_{\text{int}}}^2(\Gamma)}\right)}{\sqrt{2\pi\:{Q_{\text{int}}}^2(\Gamma)}}.
\label{simple_ou_fp_soln}
\end{equation}

Compared to \cite{avlonitis2000}, we find that the presence of an Ornstein-Uhlenbeck noise instead of a white noise, results in a scaling of the $Q_{\text{int}}$ terms as $\sqrt{\tau}$, a distinct dependence of the dynamics on the decay time.

\subsection{Generalized Ornstein-Uhlenbeck (GOU) Process}
\label{general_ornstein_uhlenbeck}

The representation of this process is $\frac{F(u)}{G(u)}=u^n$; $n>1$. As a typical case, we will consider $n=2$ which will be defined as the next higher-ordered term to the standard Ornstein-Uhlenbeck $\frac{F(u)}{G(u)}=u$ case considered in case B above. 

In this case, the relevant Fokker-Planck equation can be shown to be

\begin{eqnarray}
\frac{\partial P}{\partial \Gamma}&=& -\frac{\partial}{\partial u}\big(u^2 P\big)+\:\frac{{Q_0}^2}{2} \bigg[ (1+2\tau u)\frac{\partial^2 P}{\partial u^2} \nonumber \\
&+& 4\tau \frac{\partial P}{\partial u} \bigg].
\label{gen_ou_fp_Equation}
\end{eqnarray}

\begin{center}
\begin{figure}[tbp]
\includegraphics[height=8.0cm,width=9cm]{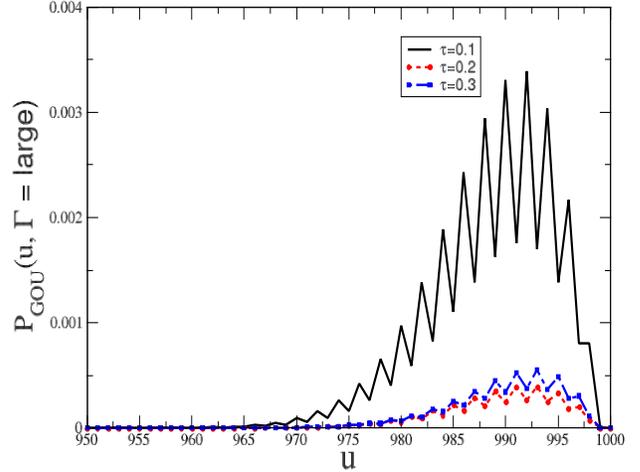}
\caption{Plots of GOU-PDFs versus the variable $u$, solved from Eq. (\ref{gen_ou_fp_Equation}), for large $\Gamma$ (steady state limit) and for different values of the decay time $\tau$.
\label{fig_genou}}
\end{figure}
\end{center}

The steady state form of Eq. (\ref{gen_ou_fp_Equation}), defined through the identity $\frac{\partial P}{\partial \Gamma}=0$, admits of the following solution:

\begin{eqnarray}
P(u,\Gamma \to \infty) &=& e^{-\frac{(1-u\tau)u}{2\tau^2 {Q_0}^2}}\:{(1+2\tau u)}^{-1+\frac{1}{4\tau^3 {Q_0}^2}} \nonumber \\
&\times& \bigg[-1-2\tau e^{-\frac{1}{16\tau^3}}\:\sqrt{\pi \tau}\:\bigg(\text{Erfi}\bigg(\frac{1}{4\tau^{3/2}}\bigg)\nonumber \\
&+& \text{Erfi}\bigg(\frac{-1+2\tau}{4\tau^{3/2}}\bigg)\bigg)\bigg],
\label{steady_state_ou}
\end{eqnarray}
\noindent
where $\text{Erfi}(z)=\dfrac{2}{\sqrt{\pi}}\:\displaystyle \int_z^\infty e^{-t^2}\:dt$.
In the asymptotic limit of fast relaxation, that is $\tau \to 0$, up to leading order, 

\begin{equation}
{P(u,\Gamma \to \infty)|}_{\tau \to 0} \sim \tau^3.
\label{ou_scaling}
\end{equation}

Clearly, the order of the polynomial has a strong influence on the relaxation rate of the system; in other words, higher ordered Ornstein-Uhlenbeck processes get increasingly affected by the noise relaxation rate. Generally, if $u^n$ represents the form of the Ornstein-Uhlenbeck process, the $\tau$-dependence of the steady-state PDF scales as $\tau^{2n+1}$.

\begin{center}
\begin{figure}[tbp]
\includegraphics[height=8.0cm,width=9cm]{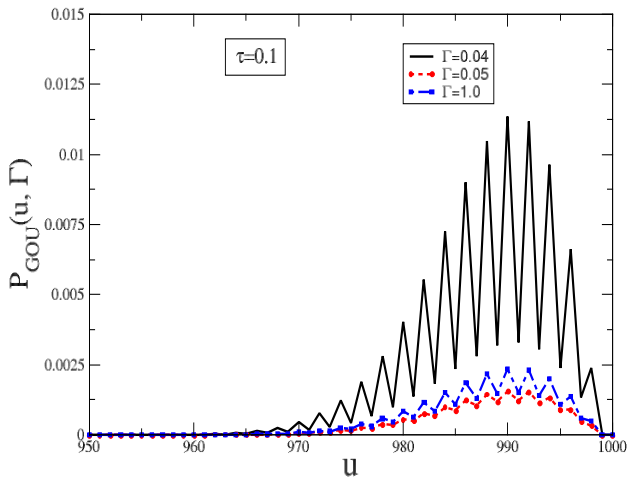}
\caption{Plots of GOU-PDFs versus the variable $u$, solved from Eq. (\ref{gen_ou_fp_Equation}), for different time regimes represented by varying $\Gamma$s for the decay time $\tau=0.1$.
\label{fig_genoutau01}}
\end{figure}
\end{center}

In order to obtain more detailed insights in to the probability distribution dynamics, we resorted to a 
numerical solution of Eq. (\ref{gen_ou_fp_Equation}). The results were tested independently using an Euler discretization scheme as well as a Runge-Kutta (order 4) with both numerical schemes providing identical results. In order to conform and compare with the general Gaussian noise case as studied in \cite{avlonitis2000}, we used $Q_0=2$ (high noise level) and $\Gamma$-values within the same range as in \cite{avlonitis2000}. The results plotted below were for simulations that ran over 1000 space points and over 10,000 time steps. We have also studied larger spatiotemporal domains but apart from numerical difference (and numerical time cost), all conclusions remained unchanged.

\begin{center}
\begin{figure}[tbp]
\includegraphics[height=8.0cm,width=9cm]{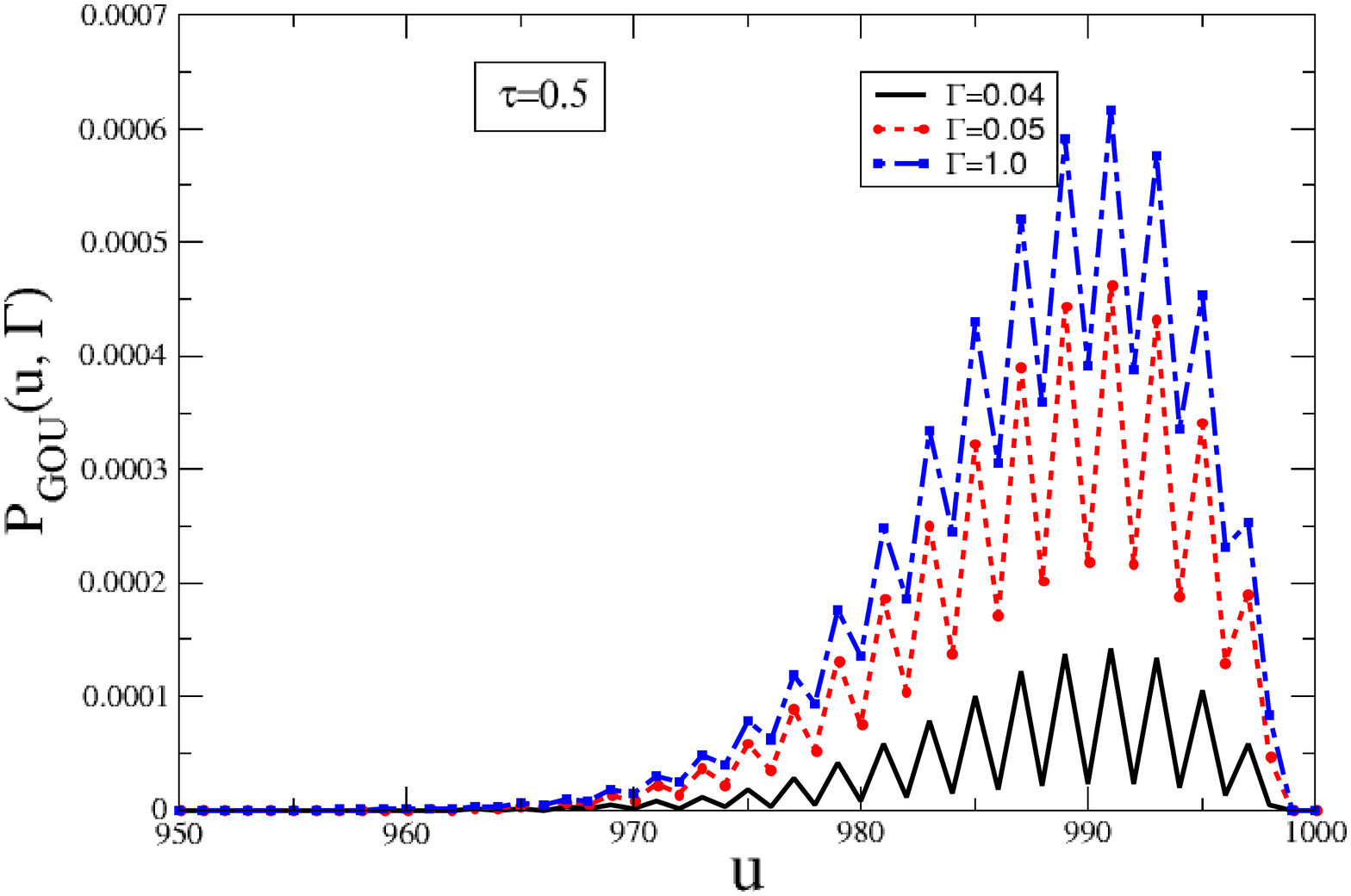}
\caption{Plots of GOU-PDFs versus the variable $u$, solved from Eq. (\ref{gen_ou_fp_Equation}), for different time regimes represented by varying $\Gamma$s for the decay time $\tau=0.5$.
\label{fig_genoutau05}}
\end{figure}
\end{center}

Figure \ref{fig_genou} shows the steady state variation of the generalized Ornstein-Uhlenbeck (GOU) PDF against the variable $u$, for a range of decay times. The plots clearly show that larger the decay time, greater is the probability of \enquote{localization} of the internal variables around the maximally probable value of $u$. This can be easily comprehended from Eq. (\ref{noise_Equation}) that suggests a proximity to a normal distribution for large values of $\tau$.
The importance of the Ornstein-Uhlenbeck decay time assumes further prominence for non-steady state properties, as is evident from Figures \ref{fig_genoutau01} and \ref{fig_genoutau05}. As the decay time increases from $\tau=0.1$ to $\tau=0.5$, thereby approaching the equilibrium steady state limit, the probability of the small-$\Gamma$ (equivalent to short time regimes) modes are seen to sharply decrease as could be seen from a comparison of Figures \ref{fig_genoutau01} and \ref{fig_genoutau05}. 

The plots above show the importance of the noise distribution in that they show time dynamics that have hitherto remained inaccessible within the Gaussian noise regime. While large vales of $\Gamma$ showed almost a solitonic dynamics in  \cite{avlonitis2000}, our results allay such simplistic conclusions in favour of a far more involved dynamics involving both $\tau$ and $\Gamma$, with a clear signature of \enquote{crossover dynamics} for varying $\tau$ in the $u-\Gamma$ phase space. Generally, most of the plots show oscillatory profiles that is simply an artefact of the finite time discretization used; for all practical purposes, the envelopes of the corresponding trajectories indicate the actual profile.

\subsection{Generalized Exponential (GE) Process}
\label{general_exponential}

The representation of this process is $\frac{F(u)}{G(u)}=\exp(\alpha u)$ ($\alpha<0$); where $1/\alpha$ represents the noise decay rate. The quantity $\alpha$ is intrinsically negative and could represent a wide variety of mesoscopic material properties ranging from strain hardening, to plastic straining to slip-channel formation \cite{avlonitis2000}, thereby adding a second length of time scale to the existing structure.
In this case, the relevant Fokker-Planck equation is given by the following

\begin{eqnarray}
\frac{\partial P}{\partial \Gamma} &=&  \frac{{Q_0}^2}{2}(1+2\tau e^{\alpha u})\:\frac{\partial^2 P}{\partial u^2} + ({Q_0}^2\alpha^2 \tau-1)e^{\alpha u} \frac{\partial P}{\partial u} \nonumber \\
&+& \left(\frac{{Q_0}^2}{2}\alpha^2 \tau -1\right)\alpha e^{\alpha u}P.
\label{gen_exp_Equation}
\end{eqnarray}

\begin{center}
\begin{figure}[tbp]
\includegraphics[height=8.0cm,width=9cm]{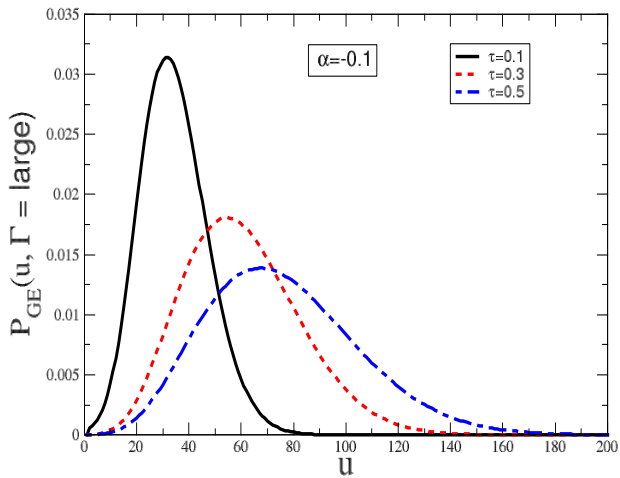}
\caption{Plots of GE-PDFs versus the variable $u$, solved from Eq. (\ref{gen_exp_Equation}), for large $\Gamma$ (steady state limit) and for different values of the decay time $\tau$, for a slowly decaying exponential function represented by the value $\alpha=-0.1$.
\label{fig_genexptau01}}
\end{figure}
\end{center}

\begin{center}
\begin{figure}[tbp]
\includegraphics[height=8.0cm,width=9cm]{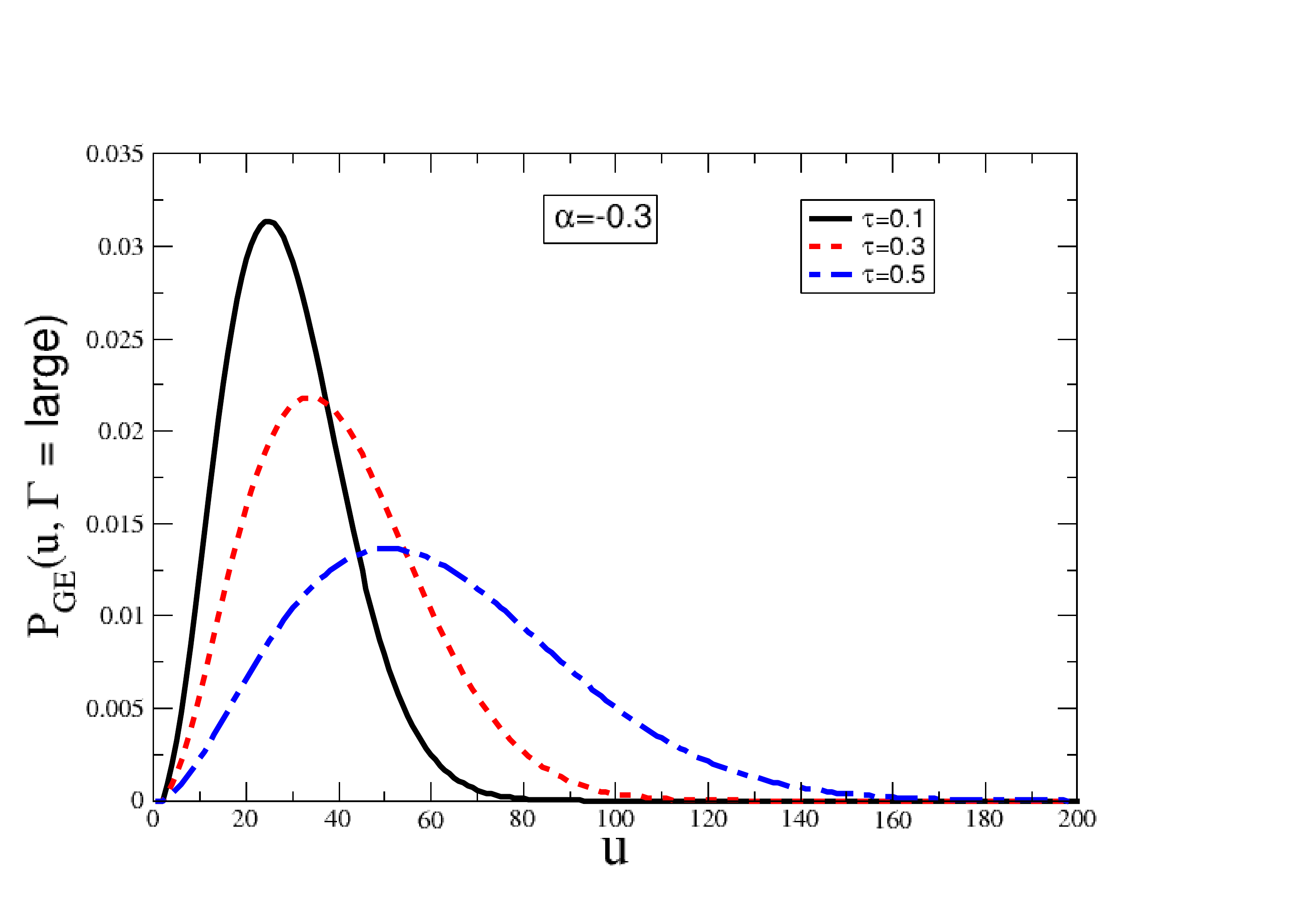}
\caption{Plots of GE-PDFs versus the variable $u$, solved from Eq. (\ref{gen_exp_Equation}), for large $\Gamma$ (steady state limit) and for different values of the decay time $\tau$, for a slowly decaying exponential function represented by the value $\alpha=-0.3$.
\label{fig_genexptau03}}
\end{figure}
\end{center}

\begin{center}
\begin{figure}[tbp]
\includegraphics[height=8.0cm,width=9cm]{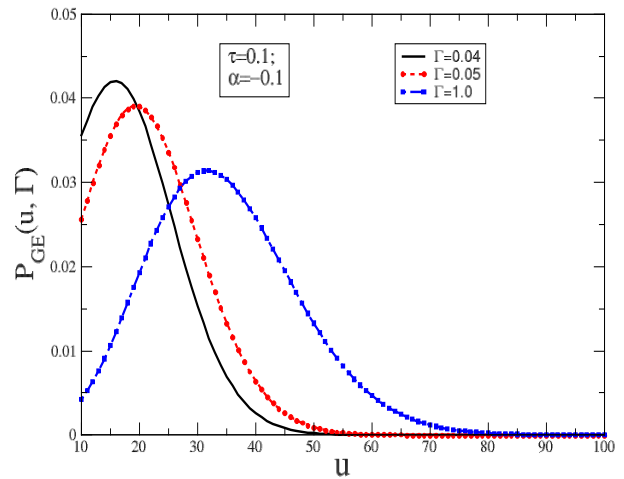}
\caption{Plots of GE-PDFs versus the variable $u$, solved from Eq. (\ref{gen_exp_Equation}), for different time regimes represented by varying $\Gamma$s for the decay time $\tau=0.1$ and the exponential decay rate $\alpha=-0.1$.
\label{fig_genexpalp01tau01}}
\end{figure}
\end{center}

Abiding the same numerical scheme as detailed in the section above, we obtained numerical solution of Eq. (\ref{gen_exp_Equation}). The primary difference between Figures \ref{fig_genexptau01} and \ref{fig_genexptau03} is the marginally lower probability for larger $\tau$ values with decrease in the value of $\alpha$ from -0.1 to -0.3. This is not difficult to perceive either; as decay time increases, a slower (exponential) relaxation rate will only decrease the probability of capturing the dislocation within the same time range. This will also lead to slower approach towards a steady state. 

In this case, the steady-state solution abides a hypergeometric form, but more importantly, for this case, the universal scaling property is lost and the steady-state PDF grows as

\begin{equation}
{P(u,\Gamma \to \infty)|}_{\tau \to 0} \sim \tau e^{\alpha u}.
\label{exp_scaling}
\end{equation}

\begin{center}
\begin{figure}[tbp]
\includegraphics[height=8.0cm,width=9cm]{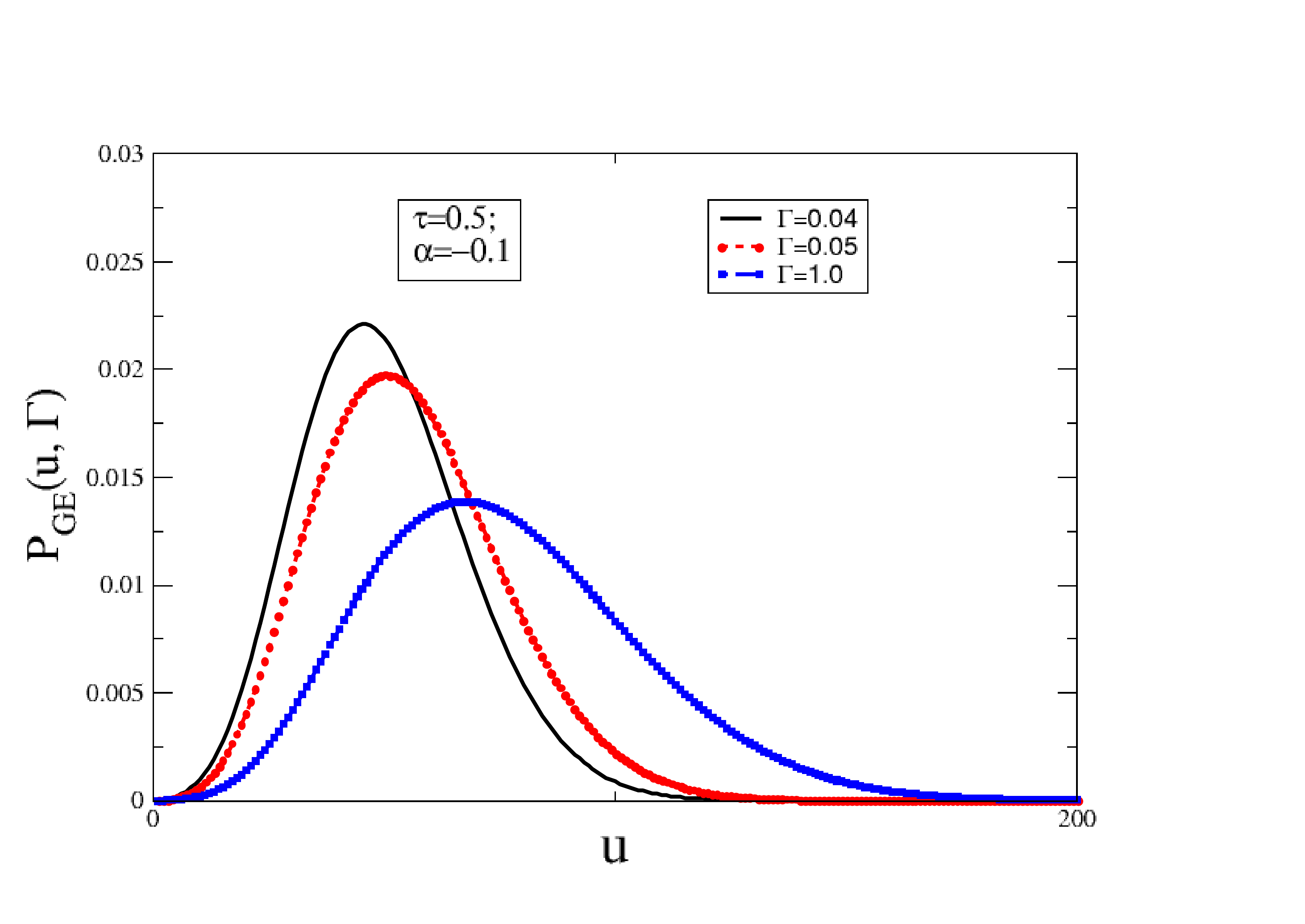}
\caption{Plots of GE-PDFs versus the variable $u$, solved from Eq. (\ref{gen_exp_Equation}), for different time regimes represented by varying $\Gamma$s for the decay time $\tau=0.5$ and the exponential decay rate $\alpha=-0.1$.
\label{fig_genexpalp01tau05}}
\end{figure}
\end{center}

A comparison of the respective plots clearly show that the generalized exponential process \enquote{localizes} at much smaller $u$-values compared to the generalised Ornstein-Uhlenbeck process. A further comparison of Figures \ref{fig_genexpalp01tau01} and \ref{fig_genexpalp01tau05} also suggests an approach towards a symmetric steady state with increasing values of the decay time $\tau$ for the same value of $\alpha=-0.1$. This is reminiscent of similar findings in \cite{avlonitis2000} which is reassuring. To avoid repetition, we refrain from adding similar plots for other values of $\alpha$ as the results have been found to remain qualitatively unchanged.  

The results presented here are expected to have vital experimental ramifications. While detailed quantitative implications of Ornstein-Uhlenbeck type decay-modified stochastic forcing on experimental designs of dislocation patterning, modified by long ranged spatial correlations, will be discussed in detail in a companion publication, the qualitative imprints can be easily found on bimodal dislocation distributions observed during cyclic deformations, as discussed in \cite{avlonitis2007, zaiser2007}. In this connection, it is also noted that the nonlinear function $F(\phi)$ in Eqn. (\ref{rescaled_Equation}) models dislocation annihilation processes as discussed within a more general framework in \cite{aifantis1982} and \cite{walgraef1985,Aifantis2006}. 

\section{Conclusion}
\label{conclusion}

While the AZA-model \cite{avlonitis2000} made the first important incision in the study of the role of stochastic perturbation in plastic deformation, the implication remained quantitatively inconclusive in absence of an appropriate noise distribution function. In choosing Ornstein-Uhlenbeck noise as our noise distribution, we have simultaneously addressed three key issues to the understanding of stochastically forced plastic deformation: a) how the presence of a finite decay time affects the approach to equilibrium, while remaining subjective of the choice of the dynamics concerned ({\it i. e.} whether the process chosen in Wiener, or Ornstein-Uhlenbeck, etc.); 2) interaction of an Ornstein-Uhlenbeck noise with an inherently Ornstein-Uhlenbeck statistical process; the results, as shown in the figures above, go way beyond simple amplification of the noisy regime, even indicating somewhat counter-intuitive slower approach to equilibrium compared to an exponential process depending on the parametric regime; and 3) hugely different relaxation modes between the generalized Ornstein-Uhlenbeck and exponential class of systems even though both have inherently exponential decay modes (admittedly, the decay mode for a generalized exponential process is spatial compared to a temporal decay for the Ornstein-Uhlenbeck process). 

An important dynamical property of note is the relatively slow growth followed by a fast decay in the GOU process while the GE process conversely shows a fast saturation to the most probable value followed by a slow decay. The four statistically plausible cases referred to in this article  - Wiener, Ornstein-Uhlenbeck, Generalized Ornstein-Uhlenbeck and Generalized Exponential - have been primarily studied for monotonically increasing dislocation densities, although a similar formulation, based on coupled dislocation dynamics, could analyze relocation-annihilation effects at the microstructure level \cite{Aifantis2006, aifantis1982}. Generalizing the scope of the present stochastic analysis to incorporate multiplicative noise could bring about such features within the framework of this model.

Our results here benchmark the implication of a finite time decaying noise spectrum in plastic deformation dynamics and thereby lays out a theoretical structure for studying a wide class of exactly solvable models perturbed by different types of noises. 

\section{Acknowledgment}
\label{acknowledgment}
The combined support of ERC-13 and ARISTEIA II projects funded by GSRT of the Green Ministry of Education, sponsoring AKC's visits to Thessaloniki, is gratefully acknowledged.

\end{document}